\documentstyle[prd,aps,epsfig,preprint,tighten]{revtex}
\def\cs{c_2}
\def\ro{{d_1}}
\def\as{\alpha_s}
\def\si{\sigma}
\def\mm{M_{\tau}^2}
\def\m{M_{\tau}}
\def\Lms{\Lambda_{\overline{\rm{MS}}}}
\def\Lmss{\Lambda_{\overline{\rm{MS}}}^2}
\newcommand{\beg}{\begin{equation}}
\newcommand{\en}{\end{equation}}
\begin{document}
\draft
\preprint{OKHEP--98--01}
\title{Analytic Perturbation Theory and\\
 Renormalization Scheme Dependence\\ in $\tau$-decay}
\author{ K. A. Milton$^a$\thanks{E-mail: milton@mail.nhn.ou.edu}\\
I. L.
Solovtsov$^{a,b}$\thanks{E-mail: solovtso@thsun1.jinr.ru} \\ V. I.
Yasnov$^b$\thanks{E-mail: yasnov@thsun1.jinr.ru}}
\address{$^a$Department of Physics and Astronomy, University of
Oklahoma, Norman, OK 73019 USA\\
{}$^b$Bogoliubov Laboratory of Theoretical Physics, Joint
Institute for Nuclear Research,\\
 Dubna, Moscow Region, 141980, Russia}
\maketitle

\begin{abstract}
We apply analytic perturbation theory in next-to-next-to-leading order
to inclusive semileptonic $\tau$-decay and study the renormalization
scheme dependence. We argue that the
renormalization  scheme ambiguity is considerably reduced
in the analytic perturbation theory framework and we obtain
a rather stable theoretical prediction.
\end{abstract}
\section{ Introduction}
The $\tau$-lepton is the only known lepton massive enough to decay
into hadrons. Its semileptonic decays are very convenient for  studying
strong interactions at low energies, because nonperturbative
contributions are rather small and one can apply the standard methods
of quantum field theory. The measurement of the quantity
$R_{\tau}$ allows one to extract the value of the strong coupling
constant at a low energy scale. 
The comparison of this value with ones obtained at
higher energies is an important  test of the applicability of QCD perturbation
theory at low energies.

The method usually used to calculate perturbative QCD contributions to 
quantities such as $R_\tau$
consists in rewriting the original expression, which involves integration over
small values of momentum, by means of Cauchy's theorem and thereby convert it
into a contour integral in a region of relatively large momentum.
But the occurrence of incorrect analytic properties of the perturbative
approximation, such as the appearance of a ghost pole and an unphysical cut, 
make it impossible to exploit Cauchy's theorem in this manner.

In this paper we will use analytic perturbation theory \cite{s3,s4}, which
provides a correlator $\Pi (q^2)$\  with the correct analytic properties.
For example, the one-loop analytic running coupling constant
$a_{\rm an}(q^2)$ in this approach is given by\footnote{We use the definition
$q^2<0$\ in the Euclidean region.}
\beg
a_{\rm an}(q^2)=\frac{2}{b}\left[\frac{1}{\ln{(-q^2/\Lambda^2)}}+
\frac{\Lambda^2}{\Lambda^2+q^2}\right],
\label{one-loop}
\en
where $a=\alpha_s/\pi$ and $b/2$ is the first $\beta$-function
coefficient.
It is clear that, unlike the standard perturbative expression, this one has no
unphysical ghost pole and, therefore, possesses the correct analytic properties,
arising from the K\"allen-Lehmann analyticity that reflects the principle
of causality.
The non\-pertur\-bative term which appears in the running coupling
constant does not change the ultraviolet limit of theory.  A distinguishing
feature of this approach is the presence of a universal limit at $q^2=0$,
independent of both the scale parameter $\Lambda$\ and choice of
renormalization scheme. This quantity turns out to be stable with respect
to higher loop corrections.

In order to compare theoretical results with experimental measurements
one has to estimate the uncertainty connected to the renormalization scheme
dependence of the perturbative approximation. It is well known that at
low energies this ambiguity becomes considerable
(see, e.g., Ref.~\cite{ree}). 
Analytic perturbation theory improves the situation and gives very
stable results over a wide range of renormalization schemes, as has been
demonstrated for the $e^+e^-$ annihilation ratio in Ref.~\cite{ee}.

We will perform the calculation of $R_{\tau}$  at the two- and three-loop 
levels and
study the renormalization scheme dependence  within analytic perturbation
theory. All results will be compared with standard perturbative ones.
This investigation is a continuation of the program initiated in Ref.~\cite{s}.

\section{Problem of QCD parametrization of $R_{\tau}$}
The ratio of hadronic to leptonic $\tau$-decay widths
\beg
R_{\tau}\equiv\frac{\Gamma(\tau^{-}\to\nu_{\tau}{\rm hadrons})}
{\Gamma(\tau^{-}\to\nu_{\tau}l^{-}{\bar{\nu}}_l)}
\en
has the following theoretical representation
\beg
R_{\tau}=3\,S_{\rm EW}(|V_{ud}|^2+|V_{us}|^2)(1+\delta_{\rm QCD}), 
\en
where $V_{ud}$ and $V_{us}$ are the CKM matrix elements,
$S_{\rm EW}$ is the electroweak factor, and $\delta_{\rm QCD}$ is the
pure QCD  correction \cite{s1}, which we will calculate perturbatively.
 It is convenient to introduce
some definitions:
 ${\rm Im}\ \Pi\sim 1+r$ for the per\-tur\-ba\-tive part of
 correlator $\Pi (q^2)$\ and $ D\sim 1+d$ for the
 per\-tur\-ba\-tive part of Adler function $ D(q^2)$.
Then one can write $\delta_{\rm QCD}$\  as an integral over timelike momentum
$s$:
\beg
\label{to-tau}
\delta_{\rm QCD}=2\int_0^{\mm}\frac{ds}{\mm}{\left(1-\frac{s}{\mm}
\right)}^2\left(1+2\frac{s}{\mm}\right) r (s).
\en
If one uses the standard perturbative approximation this integral is
ill-defined because of unphysical singularities within
the range of integration;
standard perturbation theory (PT) fails. The most useful trick
to rescue the situation is to appeal to analytic properties of the hadronic
correlator $\Pi (q^2)$ \cite{s1}. This opens up the possibility of
exploiting Cauchy's theorem by rewriting the integral in the form of a contour
integral in  the complex $q^2$-plane  with the contour being
a circle of radius $\mm$:
\beg
\label{contour} \delta_{\rm QCD}=\frac{1}{2\pi
i}\oint_{|z|=\mm}\frac{dz}{z}{\left(1-\frac{z}{\mm}\right)}^3 
\left(1+\frac{z}{\mm}\right) d (z).
\en
However, as noted, the transition to the contour representation (\ref{contour})
requires certain analytic properties of the correlator, namely, that it
must be an analytic function in the complex
\mbox{$q^2$-p}lane with a cut along the positive real axis.  The
correlator parametrized by the running coupling constant in standard
perturbation theory does not have this virtue \cite{s10,s}. Moreover, a
renormalization group analysis gives a running coupling constant
determined in the spacelike region, whereas the initial expression 
(\ref{to-tau})
for $R_{\tau}$\ contains an integration over timelike momentum. Thus, we are
in need of some method of continuing the running coupling constant
from the spacelike to the timelike region that takes into account the proper
analytic properties of the running coupling constant. Because of this
failure of analyticity Eqs.~(\ref{to-tau}) and (\ref{contour})
are not equivalent in PT.

Nevertheless, if one uses Eq.~(\ref{contour}) with the 
PT approximation for $d(z)$, thus
calculating the contour integral in a naive way, a new problem will
emerge. It is known that the lower the energy scale we consider, the greater
the uncertainty there is related to the choice of renormalization scheme (RS).
If one turns to  the $D (q^2)$-function, which can be extracted from
the ratio $R_{e^+e^-}$ for $e^+e^-$ annihilation, one finds that the
RS dependence of theoretical predictions is considerable~\cite{ree}.  At
the energy scale around $M_\tau$  this uncertainty increases
drastically. The transition to the contour integral softens this
ambiguity, but it remains large.  Moreover, as we noted, within PT
this transition to the contour integral representation cannot be performed
self-consistently.
Therefore, there is no consistent method of calculating the
inclusive decay of the $\tau$-lepton into hadrons in the framework of
standard perturbation theory.

Recently, a new approach that eliminates the two problems described above
has been proposed~{\cite{s3,s4}. This is the so-called analytic
perturbation theory (APT), in which the hadronic correlator $\Pi (q^2)$
has the correct analytic properties. In this approach, the $d_{\rm an} (q^2)$ 
function, which is analytic in the cut $q^2$-plane,
can be expressed in terms of the spectral density $\rho (\si)$, the
basic quantity in APT (to distinguish APT and PT we have introduced
subscripts ``an'' and ``pt'' instead of QCD)
\beg \label{dsr}
\label{drho} d_{\rm an}(q^2)=\frac{1}{\pi}\int^{\infty}_0\frac{d\si}{\si -
q^2} \rho (\si).
\en

We can write down some useful dispersion relations, which are
consequences of the analytic properties of the correlator. The first one is
\beg
d(q^2)=-q^2\int^{\infty}_{0} \frac{ds}{{(s-q^2)}^2}r(s),
\en
which is inverted by the following formula,
\beg r(s)=-\frac{1}{2\pi
i}\int^{s+i\epsilon}_{s-i\epsilon}\frac{dz}{z}d(z).
\en
Here, the
contour lies in the region of analyticity of the $D(q^2)$-function. Then let
us rewrite the last expression in terms of the spectral density 
$\rho (\si)$~\cite{sm}:
\beg
\label{rrho} r_{\rm
an}(s)=\frac{1}{\pi}\int^{\infty}_{s}\frac{d\si}{\si} \rho (\si).
\en
If we substitute the last formula into
integral~(\ref{to-tau}), we can easily verify that the contour representation
(\ref{contour}) holds~\cite{s}. Using the spectral representation
(\ref{drho}), we can rewrite $\delta_{\rm an}$\ in terms of $\rho (\si)$:
\beg
\label{danrho}
\delta_{\rm an}=\frac{1}{\pi}\int_0^{\infty}\frac{d\si}{\si}\rho (\si)-
\frac{1}{\pi}\int^{\mm}_0\frac{d\si}{\si}{\left(1-\frac{\si}{\mm}\right)}^3
\left(1+\frac{\si}{\mm}\right)\rho (\si).
\en
The first term in this formula is simply the value $d_{\rm an}(0)$. Once an
expression for $\rho (\si)$ is given, we can calculate the integral 
(\ref{danrho}).

Consider $\delta_{\rm an}$ at  the three-loop level.
The running coupling constant $a=\as/\pi$ satisfies the NNLO
renormalization group equation:
\beg
\mu^2\frac{da}{d\mu^2}=-\frac{b}{2}a^2(1+c_1a+\cs a^2).
\en
In the modified minimal subtraction scheme, $\overline{\rm{MS}}$, for
three active flavors, we have $b=9/2$,  $c_1=16/9$\ and
$c_2^{\overline{\rm{MS}}}=3863/864$.
The renormalization-group improved perturbation
expansion for $d_{\rm pt} (q^2)$ is given by\footnote{We
have made a few changes in notation from that given in Ref.~\cite{s}:
Now $a=\alpha_s/\pi$, and consequently we have denoted by 
$d_1$ and $d_2$  what we called
$d_2$ and $d_3$ previously, apart from a power of 4.}
\beg
d_{\rm pt} (q^2)=a(q^2)[1+\ro a(q^2)+d_2 a^2(q^2)].
\en
In the following, values of $\ro$ and $d_2$ will be needed;
in the $\overline{\rm{MS}}$-scheme they are
$d_{1}^{\overline{\rm{MS}}}=1.6398$\ and
$d_2^{\overline{\rm{MS}}}=6.3710$\ \cite{r35}.

In the APT approach, the spectral density is defined as the imaginary part of
the perturbative approximation to $d_{\rm pt} (q^2)$ on the
physical cut:
\beg
\label{e.13}
\rho (\si)=\varrho_0 (\si)+d_1 \varrho_1
(\si)+d_2\varrho_2 (\si),
\en
where
\beg \label{e.14}
\varrho_n (\si)={\rm Im}[a^{n+1}(\si+i\epsilon)].
\en
The first of these spectral densities gives
an analytic expression for the running coupling constant
 $a_{\rm an}(q^2)$:
\beg
a_{\rm an}(q^2)=\frac{1}{\pi}\int_0^\infty\frac{d\si}{\si-q^2}\varrho_0(\si).
\en
This is what was called the spacelike running coupling constant in
Ref.~\cite{s}.  In the one-loop approximation it leads to Eq.~(\ref{one-loop}).

For the following considerations, it is important that $d_{\rm an} (q^2)$\ and 
$a_{\rm an}(q^2)$\  have a universal limit
at the point $q^2=0$. This limiting value, generally, is independent of both
the scale parameter $\Lambda$ and the order of the loop expansion being
considered. Because $d_{\rm an}(0)$\ and $a_{\rm an}(0)$\  are equal to the 
reciprocal of the first coefficient of the QCD $\beta$-function,
they are also RS invariant (we consider only gauge- and mass-independent 
RSs). The existence of this fixed point will play a decisive role
in the very weak RS dependence of our results.

The next point we should like to note is that the two approaches, APT and PT,
coincide with each other in the asymptotic region of
high energies. The procedure 
of implementing analyticity described
above does not change the ultraviolet limit of the theory, i.e., $d_{\rm an}
(q^2)$\  has the same asymptotic behavior as $q^2\rightarrow -\infty$\ as
the standard PT Adler function.

Therefore, the above gives a self-consistent framework in which we can
estimate the $R_{\tau}$-ratio and extract from the measured value
of $R_\tau$ the QCD scale parameter $\Lambda$.

\section{Renormalization scheme dependence}
In this section we will investigate the RS dependence of the APT
approach.\footnote{In the framework of the conventional approach,
this problem has been studied in Ref.~\cite{r}.}
The Adler function $d_{\rm an}$, as in PT, is parametrized by a set of
RS parameters, but only two of them are independent. There is a
RS invariant combination $\omega_2$\ \cite{r42} of RS parameters
that binds them:
\beg \omega_2=\cs+d_2-c_1\ro-\ro^2,\label{rgi}
\en
which in our case equals 5.2378.
Here, $c_1$
is RS invariant in the class of schemes  mentioned above. Thus, we can
choose $\ro$ and $\cs$ as independent variables, which define
the RS, and parametrize the QCD correction in terms of them.

To find $d_{\rm an} (q^2)$ we use Eqs.~(\ref{e.13}), (\ref{e.14}), and
(\ref{dsr}) and solve the
 transcendental equation for the running
coupling constant on the physical cut lying along the positive
real axis:
\beg
\frac{b}{2}\ln\left(\frac{-q^2}{\Lmss}\right)
-i\pi\frac{b}{2}=d_1^{\overline{\rm{MS}}}-
\ro+\frac{1}{a}+c_1\ln \left(\frac{b}{2c_1}\right)+F^{(l)}(a),
\en
where at the two-loop order
\beg F^{(2)}(a)=
c_1\ln\left(\frac{c_1a}{1+c_1a}\right),
\en
and at the three-loop order
\beg
F^{(3)}(a)=F^{(2)}(a)+ c_2\int^a_0\frac{dx}{(1+c_1x)(1+c_1x+c_2x^2)}.
\en
 These expressions hold in
any renormalization scheme.

It is necessary to extract $\Lms$ from the fit in the
$\overline{\rm{MS}}$-scheme and after this procedure to pass on to
other renormalization schemes with the RS-invariant parameter
$\Lms$.  To do the fit we took the values $V_{ud}=0.9752$, $V_{us}=0.2205$,
$S_{\rm EW}=1.0194$\ and $\m=1.777$\ GeV \cite{r28,r27}.
For numerical estimations, we used the world average value
$R_{\tau}=3.633\pm 0.031$ \cite{s12}, which leads to the
following value of the QCD scale parameter in three-loops:  
$\Lambda_{\overline{\rm MS}}^{\rm an}=871\pm 155$\ MeV. The
corresponding value of $d_{\rm an} (\mm)$ is $0.1446\pm 0.0088$
and of $\alpha_{\rm an}(\mm)$ is $0.3962\pm 0.0298$. 
To compare
this result with the PT prediction  we did the same fit using the
contour integral representation (\ref{contour})
 and got  $\Lambda_{\overline{\rm MS}}^{\rm pt}
=385\pm 27$ MeV, $d_{\rm pt} (\mm)=0.1339\pm 0.0069 $, and $\alpha_{\rm pt}
(\mm)=0.3371\pm 0.0141$.
Thus, the values of the $D$-function
are close to each other, but the value of 
$\Lambda_{\overline{\rm MS}}^{\rm an}$
in the APT approach is
much larger than the one obtained in PT.  As in Ref.\ \cite{s},
we found that this value is
 very sensitive to the value of $R_{\tau}$.  In other words,
$\delta_{\rm an}$ depends so slightly on $\Lms$ that a $0.9\%$ error in 
$R_{\tau}$ gives $18\%$\ error in the value of $\Lms$.
We illustrate this feature in Table 1.
According to the table, when we change $\mm/\Lambda^2$ from 2.0 to 6.5
(corresponding to a variation of $\Lambda$ from 1.256 GeV to 0.697~GeV),
$\delta_{\rm an}$ is only altered by about $20\%$. The sensitivity to
$\Lms$\ increases as $\mm/\Lambda^2$ gets smaller.

For comparison, we also include the two-loop result: 
$\Lambda_{\overline{\rm MS}}^{\rm an}=918\pm 151$~MeV, $d_{\rm an}
(\mm)=0.1431\pm 0.0084$, and the running coupling constant
$\alpha_{\rm an}(\mm)=0.3983\pm 0.0236$. The perturbative result is
$\Lambda_{\overline{\rm MS}}^{\rm pt}=458\pm 31$ MeV,
$d_{\rm pt}(\mm)=0.140\pm 0.0067$, and 
$\alpha_{\rm pt}(\mm)=0.3544\pm 0.0157$.

The reader should further note that the convergence 
properties of the APT series are much better
than those of the PT case; in the improved scheme,
the main contribution comes from the diagrams of lowest order.
For example, the value of the three-loop APT correction to 
$d_{\rm an}(M_\tau^2)$ is smaller
than the corresponding PT contribution to $d_{\rm pt}(M_\tau^2)$
by about a factor of four.

Having found $\Lms$,
we will now study how $\delta_{\rm an}$ varies with a
change of renormalization scheme. 
A natural way to do so is to supplement results in a
certain scheme with an estimate of variability of predictions over a whole
range of  {\it a priori\/}  acceptable schemes. To restrict this
uncertainty one has to impose some requirement eliminating schemes with
unnaturally large coefficients that would introduce large
cancellations. It was proposed to calculate theoretical predictions over
the set of schemes satisfying the condition \cite{r59}
\beg
\label{domain}
|\cs | +|d_2|+c_1|\ro|+\ro^2\le 2|\omega_2|,
\en
which allows schemes with a degree of cancellation the same or smaller
than what occurs in the scheme obeying the so-called principle of minimal
sensitivity (PMS).

In the $\overline{\rm{MS}}$-scheme we adopt $\delta_{\rm an,pt}=0.1881$.
Let us consider some points belonging to the domain described above and
compare corresponding values of $\delta_{\rm pt}$. 
For example, suppose we choose the points
$A=\left(-1.6183,0\right)$
 and
$B=\left( 0.9575,0\right)$,
where the first
coordinate is $\ro$\ and the second is $\cs$. Both points lie on
the boundary of the domain, i.e., they have the same degree of
cancellation as does the PMS scheme. But $\delta_{\rm pt} (A)=0.1999$, while
$\delta_{\rm pt} (B)=0.1886$. Therefore, even with such a restriction imposed
on RSs we have about a $5\%$ deviation of PT results from each
other. The difference between APT results is much smaller: $\delta_{\rm an}
(A)=0.1866$\ and $\delta_{\rm an} (B)=0.1881$, so we have only $0.8\%$\
deviation. For clarity we display our three-loop results  in
the form of a contour  plot, in Fig.~\ref{fig1}.
Two-loop results are $\delta_{\rm pt}
(A)=0.2182$ and $\delta_{\rm pt} (B)=0.2070$. So we have again
about a $5\%$ variation in the PT results.  In APT one has $\delta_{\rm an}
(A)=0.1899$ and $\delta_{\rm an} (B)=0.1890$, which corresponds to only a 
$0.4\%$
variation.

 

Note that the $\overline{\rm{MS}}$-scheme itself does not
belong the domain (\ref{domain}). Therefore, it is worthwhile 
to consider some schemes lying
outside the domain. In Ref.~\cite{r} it was shown that the so-called
$V$-scheme~\cite{r61} lies very far from the domain described above and
gives so a large value of $\delta_{\rm pt}$ that it cannot be used at 
this low energy. For the $V$-scheme we have $\ro=-0.109$\ and $\cs=26.200$. 
The three-loop
PT result is $\delta_{\rm pt} =0.3115$, corresponding to about 
a $66\%$\ deviation from
the $\overline{\rm{MS}}$-scheme. 
On the other hand, if we turn to APT we have
$\delta_{\rm an}=0.1878$, i.e., only about a $0.2\%$ deviation from
the $\overline{\rm{MS}}$-scheme. So the $V$-scheme 
is still useful at this energy in APT.

This stability of the APT method is a consequence of the existence of the
 RS-invariant
fixed  point, $d_{\rm an} (0)$. In PT at high energies the weak RS dependence
is a consequence of the small value of the coupling constant. At lower
energies uncertainty increases. In APT, at high energies, the situation is
the same, but at  low energies the theory has a universal limit, which
restricts the RS ambiguity over a very wide range of momentum.

Another way to illustrate the remarkable stability of APT is to calculate
the spectral functions $\varrho_n (\si)$ given by Eq.~(\ref{e.14});
one sees that $\varrho_1 (\si)$ is
 much smaller than $\varrho_0 (\si)$ over the whole spectral region. 
 The same statement is true for
the relationship between $\varrho_1 (\si)$ and
$\varrho_2 (\si)$.  This monotonically decreasing
behavior reduces the RS dependence strongly, since the
perturbative coefficients $d_1$\ and $d_2$\ in expression  (\ref{e.13}) for
$\rho(\si)$ are multiplied by these functions.

\section{Conclusion}

      We have considered inclusive $\tau$-decay in three-loop order
within analytic perturbation theory, which, in contrast to  standard
perturbation theory, is a self-consistent procedure. We summarize the
following important features of this method.

First, the method maintains the correct analytic properties and leads to
a self-consistent definition of the procedure of analytic continuation.

Second, it gives an unusually large value of the QCD scale parameter
$\Lambda$, connected to the presence of nonperturbative
contributions that appear in the APT method. The value of $\Lambda$ is
very sensitive to the precise value of $R_{\tau}$.

Third, the three-loop correction is smaller than the one found in standard
perturbation theory. Hence, the APT approach is stable with respect to 
higher loop corrections.

Fourth, the RS dependence of the results obtained is reduced drastically.
For example, the $V$-scheme, which gives a very large discrepancy in
standard perturbation theory,  can be used in
analytic perturbation theory without any difficulty. Thus, the APT
predictions are practically RS independent over a wide region of RS parameters.

Therefore, we have a self-consistent method within analytic perturbation
theory of calculating semileptonic $\tau$-decay.

In this paper,
we have not considered the standard nonperturbative power corrections.
The process of enforcing analyticity
changes the perturbative contributions by incorporating
some non\-per\-turbative terms; consequently,  power corrections should also
be changed. The role of power corrections is still  not understood within
the APT approach, and requires a separate discussion.  We hope 
to accomplish this  in our subsequent papers.

\section*{Acknowledgements}

The authors would like to thank D.V.~Shirkov and O.P. Solovtsova
for interest in this 
work and for useful comments. Partial support of the work by the US 
National Science Foundation, grant PHY-9600421, and by the US 
Department of Energy, grant DE-FG-02-95ER40923, is gratefully 
acknowledged.  The work of ILS is also supported in part by the
University of Oklahoma, through its College of Arts and Science,
International Programs, Vice President for Research, and the Department
of Physics.


\begin{figure}[tbp]
\centerline{ \epsfig{file=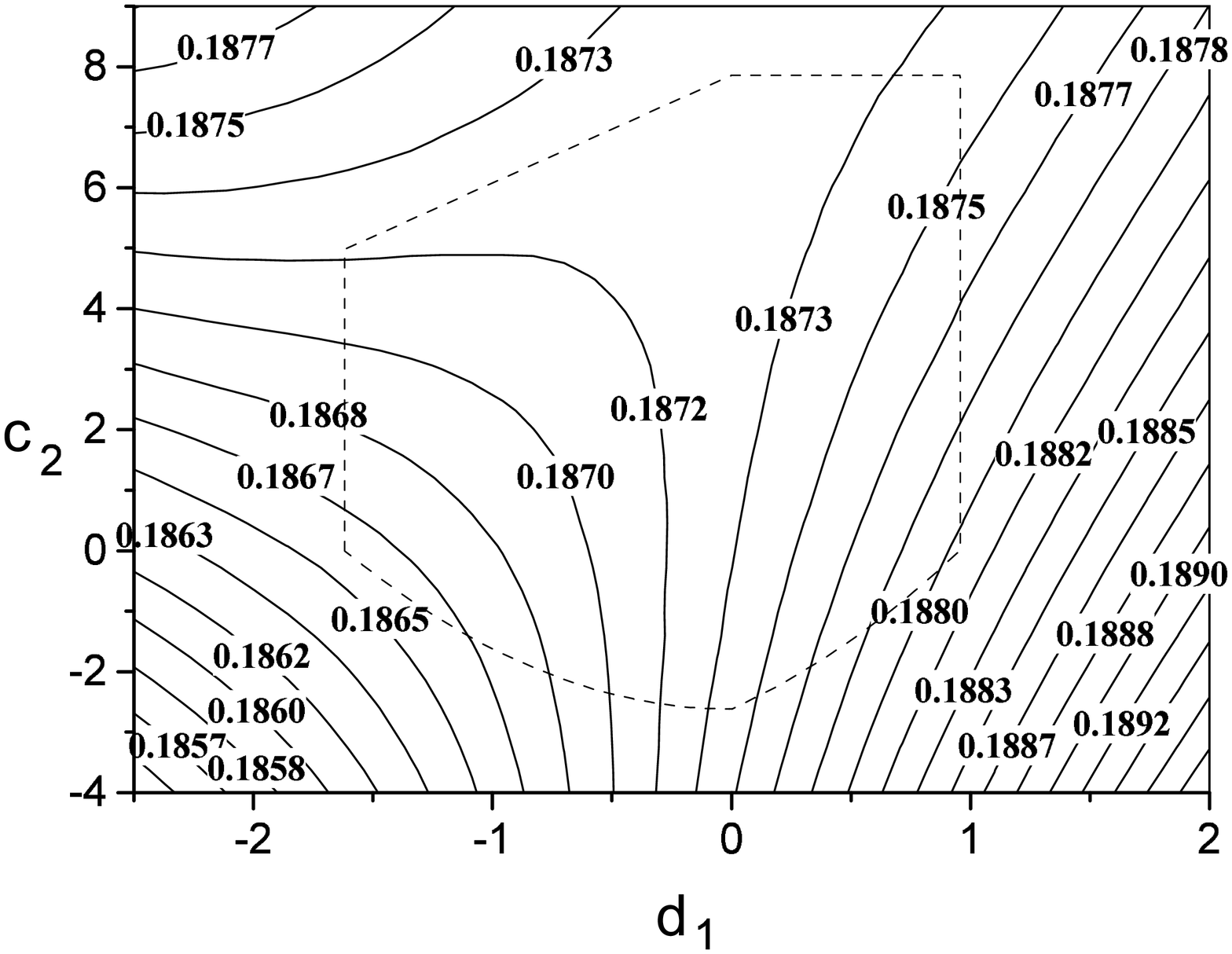,width=13.7cm}}
         \vspace{5 mm}
         \caption{ Contour plot of values of $\delta_{\rm an}$ at the three-loop
    order as a function of RS parameters $d_1$\ and $c_2$. The dashed
    line indicates the boundary of the domain, defined by
    Eq.~(\protect\ref{domain}).}
    \label{fig1}
\end{figure}

\begin{table}
\label{Tab1}
\begin{tabular}{cccccc}
$\mm/\Lambda^2$ &$\delta_{\rm an}^{{\rm NLO}}$ & 
$\delta_{\rm an}^{{\rm NNLO}}$ & $\mm/\Lambda^2$
&$\delta_{\rm an}^{{\rm NLO}}$ & $\delta_{\rm an}^{{\rm NNLO}}$ \\ \hline
2.0&0.2090&0.2106&4.5&0.1820&0.1857 \\ 
2.5&0.2016&0.2039&5.0&0.1785&0.1824 \\ 
3.0&0.1955&0.1983&5.5&0.1753&0.1795 \\ 
3.5&0.1904&0.1935&6.0&0.1724&0.1767 \\ 
4.0&0.1859&0.1894&6.5&0.1698&0.1743 \\ 
\end{tabular}
\caption{NLO and NNLO predictions for $\delta_{\rm an}$ in the
$\overline{\rm{MS}}$\ scheme. } 
\end{table}

\end{document}